\documentclass[a4paper]{jpconf}
\usepackage{graphicx}
\usepackage{slashed}
\usepackage{wrapfig}
\begin{document}
\title{Electromagnetic  neutrino:
The basic processes and astrophysical probes}

\author{Alexander Studenikin$^{1, 2}$}

\address{$^1$ Department of Theoretical Physics, Moscow State University, 119992 Moscow, Russia}
\address{$^2$ Dzhelepov Laboratory of Nuclear Problems, Joint Institute for Nuclear Research, 141980 Dubna, Russia}

\ead{studenik@srd.sinp.msu.ru}

\begin{abstract}
After a brief reminder on the electromagnetic properties of neutrinos, the main processes of the electromagnetic interactions of neutrinos in astrophysics and the corresponding limitations on millicharges and effective magnetic moments of the particle are discussed.
\end{abstract}

\section{The theory}
%\begin{wrapfigure}[16]{r}{0.5\linewidth}
\begin{wrapfigure}{r}{40mm}
%\begin{figure}[h]
\begin{center}%\begin{minipage}{14pc}
\includegraphics[width=10pc]{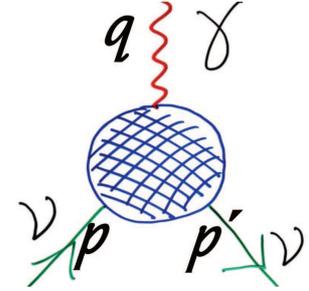}
\caption{\label{label}A neutrino effective one-photon coupling.}
%\end{minipage}\hspace{2pc}%
\end{center}
\end{wrapfigure}
%\end{figure}

The electromagnetic properties of neutrinos are embodied by the structure of the amplitude corresponding to the Feynman diagramme shown in Fig.1 . The diagramme in Fig.1 describes the electromagnetic interactions of a neutrino field $\nu (x) $ with a real photon given by the effective interaction Hamiltonian
\begin{equation}
H^{\nu}_{em} (x)= j^{\nu}_{\mu}(x) A^{\nu} (x) = {\bar \nu(x)} \Lambda _{\mu} \nu (x) A^{\mu} (x),
\end{equation}
where $j^{\nu}_{\mu}(x)$ is the neutrino effective electromagnetic current and $\Lambda_{\mu}$ is a matrix in the spinor space. Considering the neutrinos as free particles, and with the Fourier expansion of the corresponding free Dirac fields, for the amplitude corresponding to the diagramme in Fig.1 it is possible to get (see \cite{Giunti:2014ixa} and \cite{Nowakowski:2004cv} for the detailed derivations)
\begin{equation}
\langle \nu(p_f)|j^{(\nu)}_\mu (0)|\nu(p_i)\rangle=\bar{u} (p_f)\Lambda_{\mu}(q)u(p_i),
\end{equation}
where $q=p_i-p_f$. In the most general form the neutrino electromagnetic vertex function $\Lambda_{\mu}^{ij}(q)$ can be expressed \cite{Giunti:2014ixa} in terms of four form factors
\begin{equation}\label{Lambda}
\Lambda_{\mu}^{ij}(q) =  \left( \gamma_{\mu} - q_{\mu}
\slashed{q}/q^{2} \right) \left[ f_{Q}^{ij}(q^{2}) + f_{A}^{ij}(q^{2})
q^{2} \gamma_{5} \right] \nonumber
 - i \sigma_{\mu\nu} q^{\nu} \left[ f_{M}^{ij}(q^{2}) +
i f_{E}^{ij}(q^{2}) \gamma_{5} \right] ,
\end{equation}
where $\Lambda_{\mu}(q)$ and form factors $f_{Q,A,M,E}(q^2)$ are $3\times 3$ matrices in  the space of massive neutrinos.
Note that in the derivation of the decomposition (\ref{Lambda}) the demands followed from the Loretz-invariance and electromagnetic gauge invariance are used.

In the case of coupling with a real photon ($q^2=0$) the form factors $f(q^2)$ provide  four sets of neutrino electromagnetic characteristics: 1) the electric millicharges $q_{ij}=f_{Q}^{ij}(0)$,
2) the dipole magnetic moments $\mu_{ij}=f_{M}^{ij}(0)$, 3) the dipole electric moments $\epsilon_{ij}=f_{E}^{ij}(0)$ and
4) the anapole moments $a_{ij}=f_{A}^{ij}(0)$. The expression (\ref{Lambda})  for $\Lambda_{\mu}^{ij}(q)$ is applicable for Dirac and Majorana neutrinos. However, a Majorana neutrino does not have diagonal electric charge and dipole magnetic and electric form factors, only a diagonal anapole form factor can be nonzero. At the same time, a Majorana neutrino can also have nonzero off-diagonal (transition) form factors.
\begin{wrapfigure}{r}{80mm}
%\begin{figure}[h]
\begin{center}%\begin{minipage}{14pc}
\includegraphics[width=20pc]{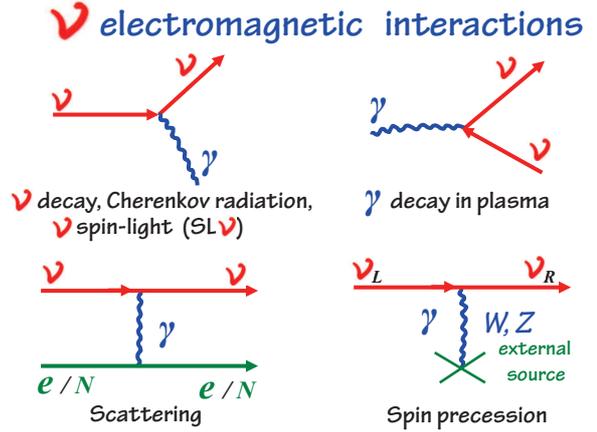}
\caption{\label{label} Basic neutrino electromagnetic processes}
%\end{minipage}\hspace{2pc}%
\end{center}
%\end{figure}
\end{wrapfigure}

For a more detailed and recent discussion on neutrino electromagnetic form factors and the reactor and solar neutrino bounds on neutrino effective magnetic moments see \cite{Studenikin_EPS_HEP_2021}.

Neutrinos with nonzero electromagnetic characteristics, due to their couplings  with real and virtual photons, generate processes that can occur in various astrophysical conditions and be the cause of important phenomena that are fundamentally observable. The most important neutrino electromagnetic processes are shown in Fig. 2 (see also \cite{Giunti:2014ixa} and \cite{Raffelt:1996wa}). They are the following: 1) a heavier neutrino decay to a lighter mass state in vacuum, 2) the Cherenkov radiation by a neutrino in matter or an external magnetic field, 3) the spin-light of neutrino in matter, 4) the plasmon decay to a neutrino-antineutrino pair in matter, 5) the neutrino scattering on an electron or a nuclei, and 6) the neutrino spin precession in an external magnetic field or the transversally moving (or the transversally polarized) matter. All of these processes can be of great interest in astrophysics, and the registration of possible consequences of these processes in experiments allows us to obtain information about the values of the electromagnetic characteristics of neutrinos and set appropriate limits. So, indeed, astrophysics can be considered as a laboratory for studying the electromagnetic properties of neutrinos (see \cite{Giunti:2014ixa}, \cite{Raffelt:1996wa} and \cite{Brdar:2020quo}).

\section{Neutrino radiative decay}

A heavier neutrino mass state $\nu_i$ in case the neutrino have nonzero electric charges (millicharges) or the magnetic and electric (the diagonal and transition) dipole moments can decay into a lighter state $\nu_f$, $m_i>m_f$, with emission of a photon. For the first time this kind of processes were discussed in \cite{Shrock:1974nd} and in the concrete applications to neutrinos in \cite{Petcov:1976ff} and \cite{Zatsepin:1978iy}. For more recent papers and detailed discussions of neutrino radiative decay $\nu_i \rightarrow \nu_f + \gamma$ see \cite{Giunti:2014ixa} and \cite{Raffelt:1996wa}. In case one neglects the effect of nonzero neutrino millicharges ($q_\nu=0$) the neutrino electromagnetic vertex function reduces to
% \begin{equation}
$\Lambda^{if}_{\mu}
=- i \sigma_{\mu\nu}  q^{\nu}
\left( \mu_{if} + i \epsilon_{if} \gamma_{5}
\right)$ ,
%\label{A209}
%\end{equation}
and
the corresponding neutrino interaction Lagrangian reads
\begin{equation}\label{Lagrang}
L=\frac{1}{2}{\bar\psi}_i\sigma_{\mu\nu}(\mu_{ij}+ i\gamma_5\epsilon_{ij})\psi_j F^{\mu\nu} + h.c.
\end{equation}
The decay rate in the rest frame of the decaying neutrino $\nu_{i}$ is given by
\begin{equation}
\Gamma_{\nu_{i}\to\nu_{f}+\gamma}
= \frac{1}{8\pi} \left( \frac{m_{i}^{2}-m_{f}^{2}}{m_{i}}
\right)^3 \left( |\mu_{fi}|^{2}
+ |\epsilon_{fi}|^{2}
\right). \label{decrat}
\end{equation}
Note that due to $m_i\neq m_f$ only the transition magnetic and electric dipole moments contribute. Therefore this expression (\ref{decrat}) is equally valid  for both Dirac and Majorana neutrinos. In the simplest extensions of the Standard Model the Dirac and Majorana neutrino decay rates are given by
\cite{Shrock:1982sc}
\begin{equation}
\Gamma_{\nu_{D}
%_{i}\to\nu^{D}_{f}+\gamma
}
\simeq
\frac{\alpha}{2}
\left(
\frac{3 G_{F}}{32 \pi^2}
\right)^2
\left(\frac{m^{2}_{i}-m^{2}_{f}}{m_{i}}\right)^3
\left(m^{2}_{i}+m^{2}_{f}\right)
\left|
U
\right|^{2} , \ \ \ \ U=\sum_{l=e,\mu,\tau}
U^{*}_{l i} U_{l f}
\frac{m_{l}^2}{m_{W}^2},
\label{decradDirac}
\end{equation}
and
\begin{equation}
\Gamma_{\nu_{M}
%_{i}\to\nu^{M}_{f}+\gamma
}\simeq
\alpha
\left(
\frac{3 G_{}}{32 \pi^2}
\right)^2
\left(\frac{m^{2}_{i}-m^{2}_{f}}{m_{i}}\right)^3
\left\{
\left(m_{i}+m_{f}\right)^2
\left|
U
\right|^{2}
\right.
\left.
- 4 m_{i} m_{f}
\left(
Re \ U
 \right)^2 \right\} .
\label{decM}
\end{equation}
Thus, for both Dirac and Majorana neutrinos we have
\begin{equation}
\Gamma_{\nu_{M}}=\frac{\mu^2_{eff}}{8\pi}\left(\frac{m^{2}_{i}-m^{2}_{f}}{m_{i}}\right)^3
\approx 5 \left(\frac{\mu_{eff}}{\mu_B}\right)^2\left(\frac{m^{2}_{i}-m^{2}_{f}}{m_{i}}\right)^3\left(\frac{m_i}{1 \ eV} \right)^3 \ s^{-1},
\end{equation}
\begin{equation}
\mu_{eff}^2= |\mu_{fi}|^{2}
+ |\epsilon_{fi}|^{2}.
\end{equation}
The corresponding life time of neutrinos in respect to the radiative decay is indeed huge
\begin{equation}
\tau_{\nu_i\rightarrow \nu_j +\gamma}\approx 0.19 \left( \frac{m^2_i}{m^2_i-m^2_j} \right)^3\left(\frac{eV}{m_i}\right)\left( \frac{\mu_B}{\mu^2_{eff}}\right) s.
\end{equation}
This is because the neutrino transition moments are suppressed by the Glashow–Iliopoulos-Maiani cancellation mechanism. Note that for the      diagonal moments there is no GIM cancellation.

The neutrino radiative decay has been constrained from the absence of decay photons in studies of the solar, supernova and reactor (anti)neutrino fluxes, as well as of the spectral distortions of the cosmic microwave background radiation. However, the corresponding upper bounds on the effective neutrino magnetic moments \cite{Raffelt:1999tx} are in general less stringent than the astrophysical bounds from the plasmon decay into neutrino-antineutrino pair discussed in the next section.

\section{Plasmon decay to neutrino-antineutrino pair}

For constraining neutrino electromagnetic properties, and obtaining upper bounds on neutrino magnetic moments in particular, the most interesting process is the plasmon decay into a neutrino-antineutrino pair \cite{Bernstein:1963qh}.
This plasmon process becomes kinematically allowed in media where the photon behaves as a particle with an effective mass $\omega_\gamma$. In the case of nonrelativistic plasma  the dispersion relation for a photon (plasmon) is $\omega^2_{\gamma}+{\vec k}^2_{\gamma}=\omega^2_{\gamma}$,  where $\omega_{\gamma}=4\pi N_e / m_e$ is the plasmon frequency. The plasmon decay rate is given by

\begin{equation}
\Gamma_{\gamma^{*}\to\nu {\bar\nu}}
=
\frac{\mu_{eff}^{2}}{24\pi}
\,
Z
\,
\frac{(\omega_{\gamma}^{2} - k_{\gamma}^{2})^{2}}{\omega_ \gamma}
,
\label{plasmondecrat}
\end{equation}
where $Z$ is a factor which depends on the polarization of the plasmon. A plasmon decay into a neutrino-antineutrino pair transfers the energy $\omega_\gamma$ to neutrinos that can freely escape from a star and thus can fasten the star cooling. The corresponding energy-loss rate per unit volume is
\begin{equation}
Q_{\gamma^{*}\to\nu {\bar\nu}}=\frac{g}{(2
\pi)^3}\int\Gamma_{\gamma^{*} \to\nu {\bar
\nu}}f_{k_{\gamma}}\omega_{\gamma}d^3k_{\gamma},
\end{equation}
where $f_{k_\gamma}$ is the photon Bose-Einstein distribution function and $g=2$ is the number of polarization states.  The magnetic moment plasmon decay enhances the Standard Model photo-neutrino cooling by photon polarization tensor, and   more fast  star  cooling slightly reduces the core temperature. These nonstandard energy losses can delay the  helium ignition in low-mass red giants. This, in turn, can be related to astronomically observable luminosity of stars before and after the helium flash. And in order not to delay the helium ignition in an unacceptable way (a significant brightness increase is constraint by observations) the upper bound on the effective neutrino magnetic was obtained in \cite{Raffelt-Clusters:90}
\begin{equation}
\mu_{eff}= \left(\sum_{fi}|\mu_{fi}|^{2}
+ |\epsilon_{fi}|^{2}\right)^{\frac{1}{2}}\leq 3\times 10^{-12} \mu_B.
\end{equation}
Recently new analysis \cite{Viaux-clusterM5:2013, Arceo-Diaz-clust-omega:2015, Capozzi:2020cbu} of the observed properties of globular cluster stars provides new upper bounds on the effective neutrino magnetic moment
%\begin{equation}
$\mu_{eff} \leq (1.2{-}2.6) \times
10^{-12} \mu _B$ that is valid for both cases of
Dirac and Majorana neutrinos.

It is interesting to compare these astrophysical bounds on the effective neutrino magnetic moments with constraints  obtained in investigations of the elastic scattering of a flavour neutrino $\nu_{l} + e^{-} \to \nu_{l} + e^{-}, \ \ l=e, \mu, \tau$ (or an antineutrino $ { \bar \nu}_l$ ) in the laboratory experiments. The bound obtained with the reactor antineutrinos by the GEMMA collaboration \cite{GEMMA:2012} is $\mu_{\nu} < 2.9 \times 10^{-11} \mu_{B}$. In investigations of the solar neutrinos the Borexino collaboration \cite{Borexino:2017fbd} obtained the bound ${\mu}_{\nu_e}< 2.8 \times
10^{-11} \mu _B$. Note that comparisons of these different experiments should be carried out with caution, since in scattering cross section is sensitive to the effective magnetic moment $\mu_{\nu_{l}}$ 
\begin{equation}
{\mu_{\nu_{l}}}^{2}(L,E_{\nu})
=
\sum_{j}
\left|
\sum_{k}
U_{l k}^{*}
e^{- i \Delta{m}^{2}_{kj} L / 2 E_{\nu} }
\left(
\mu_{jk} - i \epsilon_{jk}
\right)
\right|^{2}
\label{mueff}
\end{equation}
that depends on the neutrino flavour composition in the detector and accounts for the effect of mixing. For a deteiled discussion of this issue see \cite{Giunti:2014ixa} and \cite{Kouzakov:2017hbc}.

The plasmon decay considered in the vicinity of red giants can also be used to constrain neutrino millicharges $q_\nu$. The plasmon decay to neutrino-antineutrino pair is described by the Lagragian
\begin{equation}
L=- iq_\nu{\bar \psi}_\nu\gamma_\mu\psi_\nu A^\mu.
\end{equation}
In order to avoid the delay of helium ignition in low-mass red giants the millicharge should be constraint at the level $q_\nu < 2\times 10^{-14} e_0$, and from the absence of the anomalous energy-dependent dispersion of the SN1987A neutrino signal it should be $q_\nu < 3\times 10^{-17} e_0$ ($e_0$ is the value of an electron charge).

The most stringent astrophysical constraint on neutrino millicharges
%\begin{equation}\label{q_astr}
$q_{\nu}< 1.3 \times 10^{-19} e_0 $
%\end{equation}
 was obtained  \cite{Studenikin:2012vi} in consideration of the impact of the {\it neutrino star turning} mechanism ($\nu ST$) that can shift a magnetised pulsar rotation frequency. Note that neutrino millicharges
are strongly constrained on the level $q_{\nu}\sim 10^{-21} e_0$  from neutrality of the hydrogen atom.

\section{Neutrino spin conversion}

One of the most important  for astrophysics  consequences  of neutrino nonzero effective magnetic moments (see \cite{Giunti:2014ixa}, \cite{Raffelt:1999tx}, \cite {Studenikin:2004bu},  \cite{Pustoshny:2018jxb} and \cite{Studenikin:2020nky}) is the neutrino helicity change $\nu_L \rightarrow \nu_R$ with the appearance of nearly sterile right-handed neutrinos $\nu_R$ . In general, this phenomena can proceed in three different mechanisms: 1) the helicity change in the neutrino magnetic moment scattering   on electrons (or protons and neutrons), 2) the neutrino spin and spin-flavour precession in an external magnetic field, and 3) the neutrino spin and spin-flavour precession in the transversally moving matter currents or in the transversally polarized matter at rest \cite {Studenikin:2004bu},  \cite{Pustoshny:2018jxb}. For completeness note that the important astrophysical consequence of nonzero neutrino millicharges is the neutrino deviation from the rectilinear trajectory.

The detection of the SN 1987A neutrinos provides the energy-loss limits on the effective neutrino magnetic moments related to the observed duration of the neutrino signal (see \cite{Giunti:2014ixa} and \cite{Raffelt:1999tx}). In the magnetic scattering on electrons due to the change of helicity  $\nu_L \rightarrow \nu_R$ the proto-neutron star formed in the core-collapse SN can cool faster since $\nu_R$ are sterile and are not trapped in a core like $\nu_L$  are trapped for a few seconds. The escaping $\nu_R$ will cool the core very efficient and fast ($ \sim 1 \ s $). However, the observed $ 5-10 \ s $ pulse duration in Kamioka II and IMB experiments is in agreement with the standard model $\nu_L$   trapping and cooling of the star. From these considerations it was concluded that for the Dirac neutrinos the effective magnetic moment $\mu_D \geq 10^{-12}$ is inconsistent with the SN1987A observed cooling time.

There is another approach to constrain the neutrino magnetic moment from the data on SN1987A neutrinos related to the observed neutrino energies, \cite{Giunti:2014ixa} and \cite{Raffelt:1999tx}. The right-handed neutrinos $\nu_R$, that appear due to the helicity change in the magnetic scattering in the inner SN core, have larger energies than $\nu_L$ emitted from the neutrino sphere. Then in the magnetic moment precession process $\nu_R\rightarrow \nu_L$  the higher-energy $\nu_L$ would arrive to the detector as a signal of SN1897A. And from the absence of the anomalous high-energy neutrinos again the bound on the level $\mu_D \leq 10^{-12}$ can be settled.

The work is supported by the Interdisciplinary Scientific and Educational School of Moscow University “Fundamental and Applied Space Research” and by the Russian Foundation for Basic Research under grant No.
20-52-53022-GFEN-a.

\end{document}